\documentclass[a4paper,12pt]{article}
\usepackage{indentfirst}
\usepackage[utf8]{inputenc}
\usepackage[english]{babel}
\usepackage{amsmath,amssymb}
\usepackage{dsfont} 
\usepackage{subfigure}
\usepackage{cite}
\usepackage[unicode,linktocpage=true,plainpages=false,pdfpagelabels=false]{hyperref}
\usepackage{tikz}
\usepackage[compat=1.1.0]{tikz-feynman}
\usepackage{todonotes}
\hypersetup{
  colorlinks,
  citecolor=violet,
  linkcolor=violet,
  urlcolor=blue}

\renewcommand{\Im}{\text{Im}}
\renewcommand{\Re}{\text{Re}}

\graphicspath{Pictures}

\begin{document}
\title{Kinetic coefficients in the formalism of time-dependent Green's functions at finite temperature}
\author{Viacheslav Krivorol$^{1,}$\thanks{E-mail: v.a.krivorol@gmail.com} \and Michail Nalimov$^{1,2,}$\thanks{E-mail: m.nalimov@spbu.ru}}
\date{$^1$\it Saint-Petersburg State University, 7/9 Universitetskaya Emb.,\\
St Petersburg 199034, Russia\\
$^2$Bogoliubov Laboratory of Theoretical Physics, Joint Institute for Nuclear Research, 6 Joliot-Curie, Dubna, Moscow region, 141980, Russia \\[2ex]}
\maketitle
\begin{abstract}
We discuss the microscopical justification of  dissipation in  model nonrelativistic Fermi and Bose systems with weak local interactions above phase transitions. The dynamics of equilibrium fluctuations are considered in Keldysh~-- Schwinger framework. We show that the dissipation is related to pinch singularities of the diagram technique. 
Using Dyson~-- Schwinger equation and the two-loop approximation we define and  calculate the attenuation parameter which is related to exponentiality of Green's functions decay.  We show that the attenuation parameter is the microscopic analogue of the Onsager kinetic coefficient and it is related to attenuation in the excitation spectrum. 
\end{abstract}

\section{Introduction}

Theoretical description of dissipation phenomena and dynamics of fluctuations in quantum many-body systems is an interesting and important problem of modern theoretical and experimental physics \cite{zubarev1,zubarev2,snowmass,HeavyIon}. Physically, the nature of dissipation is related to interactions between the system and its environment. However, to fully understand this phenomenon, a more detailed theoretical consideration is required. Historically, the first attempts to describe dissipation from the first principles of quantum mechanics apparently were made by Feynman and Vernon \cite{FEYNMAN1,PATRIARCA}. The proposed approach was to couple the system (finite-dimensional in the simplest case) with a model environment, for example, with a system of harmonic oscillators \cite{weiss}. In some cases, dissipation effects in systems with a model environment can be studied by obtaining the exact solution.
Further research in this area led to the creation of \textit{open quantum systems theory} \cite{OpenQuantumIntr}. The main tool of this formalism is the so-called \textit{Lindblad equation} \cite{LindbladFirstPaper,Gorini,Lindblant}. It is a generalization of the standard Neumann equation for the density matrix to the case of Markovian dissipative and non-Hamiltonian evolution \cite{Tarasov}.
This equation is the von Neumann equation with additional terms. 
These terms model the interactions between the system and the environment.
The standard problems of this formalism are the choice of the explicit form of additional terms for various models and the complexity of the microscopic justification of this choice.

As is known, dissipation is standardly described in terms of kinetic coefficients.
The standard result on the kinetic coefficients microscopic structure are the Green~-- Kubo formulas \cite{zubarev1,zubarev2,Kubo1,Kubo2}. The correlators arising within this approach have an extremely complex structure and their calculation is usually available only within the framework of numerical simulation \cite{Mouas,Liu,Volkov}. 
The detailed derivation of the general formula for the kinetic coefficients of linear hydrodynamics in the Mori's projecting operators can be found in \cite{zubarev1,zubarev2}. The generalization of Mori's method to a nonlinear case was done in \cite{AdzhemyanKuni}, but this approach has not been widely applied yet. 
Note that the methods of the linear response theory (in particular including formulas of the Green~-- Kubo type) applied to quantum systems use the method of the temperature Green's functions analytic continuation \cite{Eliashberg}. However, in complicated situations using purely temperature-based methods seems less clear compared to the time-dependent Green's functions at finite temperature \cite{KelRev2}, which directly take into account the dynamics.

We are interested in constructing hydrodynamics from the first principles and microscopic Hamiltonian using field-theoretical methods \cite{Sieberer}. The motivation for this study is the standard problem of critical dynamics associated with the ambiguity of choosing the correct system of phenomenological hydrodynamic equations (the most well-known family of models includes $A,~B,~C\ldots$ models) that describe the relaxation of the order parameter \cite{Vasiljev2}. Sometimes the choice between phenomenological models is quite difficult, and the question arises whether it is possible to build such a model based on the microscopic picture. For example, in work \cite{ELSM} for $\lambda$-transition in Bose system it was shown from microscopic consideration\footnote{The similar analysis for spin systems on a cubic lattice was done in \cite{Maghrebi}.} that the simplest model $A$ is the correct hydrodynamic model, while phenomenological considerations pointed to a more complex $F$ model. Following the ideas of paper \cite{ELSM}, our goal is to generalize this result to the case of an arbitrary equilibrium state.

In this paper we study the structure of kinetic coefficients in quantum many-body systems and calculate them. To demonstrate the main ideas, we consider the dynamics of equilibrium fluctuations in a simplified Fermi or Bose system with weak local structureless interactions. A natural tool for constructing a perturbation theory in this case is the formalism of time-dependent Green's functions at a finite temperature. The dynamics of these Green's functions is given on the Keldysh~-- Schwinger contour.
The main ideas of this framework and the relevant literature can be found in \cite{KelRev1,KelRev2,KelRev3,KelRev4,KelRev5,KelRev6,KelRev7,KelRev8,K-B,zubarev2}.
It was found that, starting from the second order of the perturbation theory, the  so-called “pinch” singularities occur (singularities on large time scales of a special form) \cite{Eliashberg,Tanaka,Sangyong,Sangyong2,Yoshimasa,Nieves} for the time-dependent Green's functions at a finite temperature. These singularities are specific for quantum field theory\footnote{Note that the similar singularities of the Keldysh perturbation theory were noticed when describing the dynamics of relativistic particles on a curved space-time background \cite{Akhmedov1,Akhmedov2,Akhmedov3,Akhmedov4}.}.
The diagrams containing the singularities were regularized and calculated in the two-loop approximation.
It was noted that this procedure followed by “dressing” the regularization parameter according to the Dyson equation makes it possible to prove the existence of exponential in time decay of the total two-point Green's function. That was not observed at the level of propagators. That is, the presence of pinch singularities in this system leads to attenuation in the quasiparticle spectrum. In this paper, in the two-loop approximation we show that it is possible to explicitly calculate the exponential decay factor as a function of temperature and chemical potential.

The article is organized as follows. In section \ref{razdtreb} we discuss the general technique of time-dependent Green's functions at a finite temperature for the microscopic model Hamiltonian of a Fermi or Bose system with weak local interactions. In section \ref{diss} we discuss the dissipation emergence mechanism for this system using the Dyson equation. The attenuation parameter is determined and calculated in the two-loop approximation. In Appendix A we discuss some technical details of the asymptotic analysis of integrals arising from the “dressing” procedure. In Appendix B we give the first coefficients of the Taylor series expansion in terms of the frequencies of some Feynman diagrams.

\section{Time-dependent
Green’s functions at finite temperature}\label{razdtreb}
We consider non-relativistic Fermi or Bose many-body system with weak local interactions “density~--~density” type. To illustrate of our ideas, the specific form of the interaction potential is not important, but the calculation of Feynman diagrams becomes more complicated in a general situation. Thus, we work in the local interactions approximation. The generalization of our technique for arbitrary non-local potential is straightforward.  
We suppose that the system is homogeneous and isotropic in space.
The well-known standard Hamiltonian \cite{Abrikosov, A} of this system is given by\footnote{In this paper $\hbar=1$ and Boltzmann constant $k_B=1$.}
\begin{equation}
\hat{H}=\int d\mathbf{x} \, \hat{\psi}^+(\mathbf{x},t)
\bigg(-\frac{\Delta}{2m}-\mu\bigg)
\hat{\psi}(\mathbf{x},t)+g\int d\mathbf{x}~ \hat{\psi}^+(\mathbf{x},t)\hat{\psi}(\mathbf{x},t)\hat{\psi}^+(\mathbf{x},t)\hat{\psi}(\mathbf{x},t).
\end{equation}
Here $m$ is the mass of particles, $\hat{\psi}$ and $\hat{\psi}^+$ are fermion or boson field operators, $g$ is the coupling constant\footnote{It could be negative in Fermi systems.}, $\mu$ is the chemical potential.
Unless otherwise indicated, all integrals in this paper are assumed to be taken with infinite limits. The sum over the spin indices of the field operators is implied.

\begin{figure}
\centering
\begin{tikzpicture}
\draw[->, thick] (-1.5,0) -- (6,0);
\filldraw[black, very thick] (0,-2.5) circle (2pt) node[anchor=west]{$t_{in}-i\beta$} -- (0,-0.6);
\draw[<-, very thick]  (2.1,-0.6) -- (4,-0.6);
\draw[->, very thick]  (0,0.6) -- (2.2,0.6);
\draw[->, very thick]  (0,-0.6) -- (0,-1.7);
\draw[black, very thick] (0,-0.6) -- (4.1,-0.6);
\draw[black, very thick] (4.1,-0.6) .. controls (5.1,-0.6) and (5.1,0.6) .. (4.1,0.6);
\draw[black, very thick] (4.1,0.6) -- (0,0.6);
\filldraw[black] (0,0) circle (2pt) node[anchor=south]{$t_{in}$};
\filldraw[black] (4.3,0) circle (2pt) node[anchor=south]{$t_f$};
\filldraw[black] (5.8,-0.5) node[anchor=south]{$t$};
\filldraw (2.5,0.6) node[anchor=south]{$R$};
\filldraw (2.5,-0.6) node[anchor=north]{$A$};
\filldraw (0,-1.4) node[anchor=east]{$T$};
\end{tikzpicture}
\caption{Keldysh~-- Schwinger contour} \label{kontur}
\end{figure}
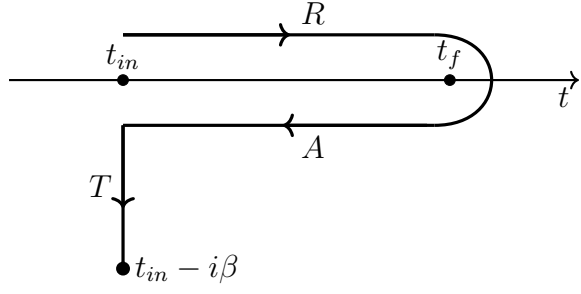

We describe the dynamics of fluctuations in the system within the formalism of time-dependent Green's functions at the finite temperature. In this framework, the standard  objects are 2-$n$ point Green’s functions
\begin{equation}
\label{GreenFunction}
G_{2n}=\mathrm{Sp}\big( \mathrm{T}\big[\hat{\psi}(\mathbf{x}_{2n},t_{2n})\ldots\hat{\psi}(\mathbf{x}_n,t_n)\hat{\psi}^+(\mathbf{x}_{n-1},t_{n-1})\ldots\hat{\psi}^+(\mathbf{x}_2,t_2) \hat{\psi}^+(\mathbf{x}_1,t_1)\big]\hat{\rho}\big),
\end{equation}
where Sp is the trace operation in the quantum mechanical sense, $\hat{\rho}$ is the density operator which describes the distribution in the system at an initial moment of time $t_{in}$.
Quantum fields are considered in the Heisenberg representation:
\begin{equation}
\hat{\psi}(\mathbf{x},t_k)=e^{i\hat{H}(t_k-t_{in})}\hat{\psi}(\mathbf{x})e^{-i\hat{H}(t_k-t_{in})},\qquad k=1,2\ldots 2n,
\end{equation}
where $\hat{\psi}(\mathbf{x})$ is an operator equal to the field operator in the Schrödinger representation at initial moment of time $t_{in}$. It is convenient not to fixing the total number of particles, thus we assume that
density matrix $\hat{\rho}$ describes the big canonical ensemble with an inverse temperature $\beta=1/T$ and the chemical potential $\mu$. We assume that the temperature $T$ and the chemical potential $\mu$ define a system without anomalous averages. Further consideration is only valid in this situation. 
The diagram technique for Green functions (\ref{GreenFunction}) is discussed in works \cite{KUFG,ELSM,osnovnaya} in terms of the operator formalism \cite{Vasiljev1}.
This framework is complicated, so we take the functional integral point of view.
The diagram technique may be obtained by considering the following generating functional \cite{N}
\begin{equation}
\int\mathcal{D}\psi^+\mathcal{D}\psi~ e^{iS+A^+\psi +A\psi^+}.
\end{equation}
Here $A$ and $A^+$ are the field sources, the symbol $\mathcal{D}$ denotes the functional integration, $\psi$ and $\psi^+$ are fermionic or bosonic fields. The action $S$ can be written in the form
\begin{align}
\label{Action}
S=\int\limits_C dt \Bigg[ \psi^+(\mathbf{x},t)\Big(i\partial_t+\frac{\Delta}{2m}+ \mu\Big)\psi(\mathbf{x},t)-g \Big(\psi^+(\mathbf{x},t)\psi(\mathbf{x},t)\psi^+(\mathbf{x},t)\psi(\mathbf{x},t)\Big) \Bigg],
\end{align}
where $C$ is the Keldysh~-- Schwinger contour \cite{K,S,Markku22} (see Figure \ref{kontur}).
The form of the action (\ref{Action}) reflects the fact that there is a set of fields $\psi_R,~\psi_A,~\psi_T$ (and the conjugates), indices $R$ (for retarded), $A$ (for advanced), $T$ (for temperature) indicate the straight-line section of the contour on which the corresponding fields are defined.
The following boundary conditions are required at the ends of the contour
\begin{equation}
\label{granusloviya}
\psi_R(t_f)=\psi_A(t_f), \qquad \psi_A(t_{in})=\psi_T(t_{in}), \qquad \psi_R(t_{in})=\pm\psi_T(t_{in}-i\beta).
\end{equation}
Hereinafter we use the shorthand notation $\pm$ and $\mp$, where sign in the upper position is used for bosons and sign in the lower position is used for fermions.
The first two conditions (\ref{granusloviya}) are the continuity conditions of the fields on the Keldysh~-- Schwinger contour.
The third condition is the standard anti-symmetry of fermionic fields and symmetry of bosonic fields similar to the one in the theory of temperature Green functions \cite{Vasiljev1}.

By $\langle\ldots\rangle_0$ we denote the averaging of fields with gaussian weight $\exp{(-S_0)}$. Here $S_0$ is the part of the action (\ref{Action}), quadratic in fields, and multiplied by $-i$.
The propagators $$\langle\psi^{~}_i(\mathbf{x},t)\psi^+_j(\mathbf{x}^\prime,t^\prime)\rangle_0\equiv G_{ij^+}(\mathbf{x},t,\mathbf{x}^\prime,t^\prime)$$  are defined as the solutions of the matrix equation \cite{Vasiljev1}
\begin{equation}
\label{PropEq}
\bigg(\partial_t-i\frac{\Delta}{2m}-i\mu\bigg)
\begin{pmatrix}
  \langle{\psi}^{~}_R{\psi}^+_R\rangle_0& \langle{\psi}^{~}_R{\psi}^+_A\rangle_0 &\langle{\psi}^{~}_R{\psi}^+_T\rangle_0\\
  \langle{\psi}^{~}_A{\psi}^+_R\rangle_0& \langle{\psi}^{~}_A{\psi}^+_A\rangle_0 & \langle{\psi}^{~}_A{\psi}^+_T\rangle_0\\
   \langle{\psi}^{~}_T{\psi}^+_R\rangle_0& \langle{\psi}^{~}_T{\psi}^+_A\rangle_0 &\langle{\psi}^{~}_T{\psi}^+_T\rangle_0\\
\end{pmatrix}
=\delta(t-t')\delta(\mathbf{x}-\mathbf{x}^\prime)\begin{pmatrix}
1 & 0 & 0\\
0 & -1 & 0\\
0 & 0 & 1
\end{pmatrix}
\end{equation}
which is consistent with the boundary conditions  (\ref{granusloviya}).
The propagators carry the indices $$i,j\in\{R,~A,~T\}$$ of the corresponding fields. Solving (\ref{PropEq}) in time~-- momentum space,
we obtain 
\begin{align}
& G_{RR^+}=\big(\theta(t-t^\prime)\pm n(\varepsilon)\big)e^{-i\varepsilon(t-t^\prime)}, & &G_{RA^+}=\pm n(\varepsilon)e^{-i\varepsilon(t-t^\prime)}, & &G_{RT^+}=\pm n(\varepsilon)e^{-i\varepsilon(t-t^\prime)},\\
& G_{AA^+}=\big(-\theta(t^\prime-t)\mp  n(-\varepsilon)\big)e^{-i\varepsilon(t-t^\prime)}, & &G_{AR^+}=\mp n(-\varepsilon)e^{-i\varepsilon(t-t^\prime)}, & &G_{AT^+}=\pm n(\varepsilon)e^{-i\varepsilon(t-t^\prime)},\\
& G_{TT^+}=\big(\theta(t-t^\prime)\pm n(\varepsilon)\big)e^{-i\varepsilon(t-t^\prime)}, & &G_{TR^+}=\mp n(-\varepsilon)e^{-i\varepsilon(t-t^\prime)}, & &G_{TA^+}=\mp n(-\varepsilon)e^{-i\varepsilon(t-t^\prime)}.
\end{align}
where $\theta(t-t^\prime)$ is the Heaviside step function, $n(\varepsilon)=(e^{\beta\varepsilon}\mp 1)^{-1}$ is the average occupation number of the level with energy $\varepsilon=\mathbf{p}^2/2m-\mu$, $\mathbf{p}$ is the momentum (with respect to $\mathbf{x}-\mathbf{x}^\prime$).

\begin{figure}[h]
\centering
\begin{tikzpicture}
\draw[->, thick] (-1.5,0) -- (6,0);
\draw[black, very thick] (-1.5,-0.6) -- (6,-0.6);
\draw[<-, very thick]  (2.1,-0.6) -- (6,-0.6);
\draw[->, very thick]  (-1.5,0.6) -- (2.2,0.6);
\draw[black, very thick] (6,0.6) -- (-1.5,0.6);
\filldraw[black] (0,0);
\filldraw[black] (4.3,0);
\filldraw[black] (5.8,-0.5) node[anchor=south]{$t$};
\filldraw (2.5,0.6) node[anchor=south]{$R$};
\filldraw (2.5,-0.6) node[anchor=north]{$A$};
\end{tikzpicture}
\caption{The Keldysh~-- Schwinger contour in the limit of $t_{in}\rightarrow-\infty,~ t_f\rightarrow+\infty$.} \label{kontur2}
\end{figure}
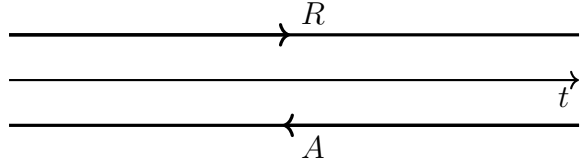

Following Keldysh method and using the freedom of choice of time moments $t_{in}$ and $t_{f}$ it is convenient to set the limits $t_{in}\rightarrow-\infty,~ t_f\rightarrow+\infty$. It is customary to facilitate the use of
the Fourier transform.
After passing the limit the system becomes homogeneous in time (see Figure \ref{kontur2}).
It is also consistent with the logic of the hydrodynamic consideration in the sense that we are interested in the limit of large time scales. The initial equilibrium distribution turns out to be given in the infinitely distant past.
Due to the specific form of the propagators and Keldysh limit, the perturbation theory  contains specific singularities of the “pinch” type \cite{Hua}  for the models similar to (\ref{Action})\cite{ELSM, N}.
To discuss them, it is
convenient to switch to the new fields called \textit{Keldysh variables}\cite{K}:
\begin{equation}
\xi=\frac{\psi_R+\psi_A}{\sqrt{2}},
\qquad
\eta=\frac{\psi_R-\psi_A}{\sqrt{2}},
\qquad
\xi^+=\frac{\psi^+_R+\psi^+_A}{\sqrt{2}},
\qquad
\eta^+=\frac{\psi^+_R-\psi^+_A}{\sqrt{2}}.
\end{equation}
In these variables the propagators are significantly simplified: 
\begin{align}
& G_{\eta\xi^+}=-e^{-i\varepsilon(t-t')}\theta(t'-t), & &G_{\xi\eta^+}=e^{-i\varepsilon(t-t')}\theta(t-t'), & &G_{\xi\xi^+}=e^{-i\varepsilon(t-t')}\big(1\pm 2n(\varepsilon)\big),\label{PropKel3}\\
& G_{\eta\eta^+}=0,\label{PropKel4} & &G_{\xi T^+}=\pm \sqrt{2}n(\varepsilon)e^{-i\varepsilon(t-t^\prime)}, & &G_{\eta T^+}=0,\\
& G_{T\xi^+}=\mp \sqrt{2}n(-\varepsilon)e^{-i\varepsilon(t-t^\prime)}, & &G_{T\eta^+}=0.
\end{align}
Let the interaction $S_{int}$ be the part of the original action, non-quadratic in fields.
In terms of the Keldysh variables the interaction can be written in the form \cite{N}
\begin{equation}
\label{SintKeld}
S_{int}=g\big(\xi^+\xi\xi^+\eta+
\xi^+\xi\eta^+\xi+
\xi^+\eta\eta^+\eta+
\eta^+\xi\eta^+\eta\big).
\end{equation}
Introduce the following graphic notation
\begin{equation}
\label{GrafOb}
G_{\eta\xi^+}=\feynmandiagram [horizontal=a to b] {a -- [insertion=0.8,anti fermion={[size=3pt]}0.3] b};,\qquad G_{\xi\eta^+}=\feynmandiagram [horizontal=a to b] {a -- [insertion=0.8, fermion={[size=3pt]}0.3] b};,\qquad G_{\xi\xi^+}=\feynmandiagram [horizontal=a to b] {a -- [insertion=0.8] b};.
\end{equation}
Taking into account the graphic notation (\ref{GrafOb}), the interaction structure (\ref{SintKeld}) leads to
the presence of the following vertices in the diagram technique:
\begin{equation}
\raisebox{-0.1em}{\begin{tikzpicture}
\begin{feynman}
\vertex (a1) ;
\vertex[left=0.7cm of a1] (a2);
\vertex[above=0.7cm of a2] (a3);
\vertex[right=0.7cm of a1] (a4);
\vertex[above=0.7cm of a4] (a5);
\vertex[left=0.7cm of a1] (a6);
\vertex[below=0.7cm of a6] (a7);
\vertex[right=0.7cm of a1] (a8);
\vertex[below=0.7cm of a8] (a9);
\diagram* {
(a3) -- [insertion=0.6](a1), 
(a5) -- (a1),
(a7) -- [insertion=0.6](a1),
(a9) -- [fermion](a1),
};
\end{feynman}
\end{tikzpicture}},\qquad\qquad
\raisebox{-0.1em}{\begin{tikzpicture}
\begin{feynman}
\vertex (a1) ;
\vertex[left=0.7cm of a1] (a2);
\vertex[above=0.7cm of a2] (a3);
\vertex[right=0.7cm of a1] (a4);
\vertex[above=0.7cm of a4] (a5);
\vertex[left=0.7cm of a1] (a6);
\vertex[below=0.7cm of a6] (a7);
\vertex[right=0.7cm of a1] (a8);
\vertex[below=0.7cm of a8] (a9);
\diagram* {
(a3) -- [insertion=0.6](a1), 
(a5) -- (a1),
(a7) -- [insertion=0.8, fermion](a1),
(a9) -- (a1),
};
\end{feynman}
\end{tikzpicture}},\qquad\qquad
\raisebox{-0.1em}{\begin{tikzpicture}
\begin{feynman}
\vertex (a1) ;
\vertex[left=0.7cm of a1] (a2);
\vertex[above=0.7cm of a2] (a3);
\vertex[right=0.7cm of a1] (a4);
\vertex[above=0.7cm of a4] (a5);
\vertex[left=0.7cm of a1] (a6);
\vertex[below=0.7cm of a6] (a7);
\vertex[right=0.7cm of a1] (a8);
\vertex[below=0.7cm of a8] (a9);
\diagram* {
(a3) -- [insertion=0.6] (a1), 
(a5) -- [fermion] (a1),
(a7) -- [insertion=0.8, fermion] (a1),
(a9) -- [fermion] (a1),
};
\end{feynman}
\end{tikzpicture}},\qquad\qquad
\raisebox{-0.1em}{\begin{tikzpicture}
\begin{feynman}
\vertex (a1) ;
\vertex[left=0.7cm of a1] (a2);
\vertex[above=0.7cm of a2] (a3);
\vertex[right=0.7cm of a1] (a4);
\vertex[above=0.7cm of a4] (a5);
\vertex[left=0.7cm of a1] (a6);
\vertex[below=0.7cm of a6] (a7);
\vertex[right=0.7cm of a1] (a8);
\vertex[below=0.7cm of a8] (a9);
\diagram* {
(a3) -- [insertion=0.75,fermion] (a1), 
(a5) -- (a1),
(a7) -- [insertion=0.8, fermion] (a1),
(a9) -- [fermion] (a1),
};
\end{feynman}
\end{tikzpicture}}.
\end{equation}
Notice that any diagrams with closed arrow cycles vanish\footnote{It follows from the retarded and the advanced structure of the propagators $G_{\eta\xi^+}$ and $G_{\xi\eta^+}$.}, similar to the Wyld diagram technique \cite{Wyld}.

\section{Pinch singularities and dissipation}\label{diss}

It can be shown that the diagram technique discussed above contains specific pinch singularities \cite{Eliashberg,Tanaka,Sangyong,Sangyong2,Yoshimasa,Nieves}. For example, consider the following diagram in time~-- momentum representation:
\begin{align}
&\raisebox{-1.4em}{\begin{tikzpicture}
\begin{feynman}
\vertex (a1) ;
\vertex[right=0.9cm of a1] (a2);
\vertex[right=0.9cm of a2] (a3);
\vertex[right=0.9cm of a3] (a4);
\vertex[right=0.9cm of a4] (a5);
\vertex[below=0.05cm of a5]
(a6);
\vertex[right=0.05cm of a6]
(a7);
\diagram* {
(a4) -- (a5), 
(a4) -- (a3),
(a4) -- [insertion=0.8,anti fermion={[size=1t]}10](a2),
(a2) -- [anti fermion,insertion=0.15](a1),
(a4) -- [out=-90, in=-90, looseness=1,insertion=0.2] (a2),
(a4) -- [out=90, in=90, looseness=1,insertion=0.2] (a2)
};
\end{feynman}
\end{tikzpicture}}=\frac{\mp \zeta g^2}{(2\pi)^6}\int d\mathbf{k} d\mathbf{q}~ e^{-i\varepsilon(\mathbf{k}) (t-t^\prime)}\big(1\pm 2n(\mathbf{k})\big) e^{-i\varepsilon(\mathbf{q}) (t-t^\prime)}\times\notag\\ &\times \big(1\pm 2n(\mathbf{q})\big)e^{i\varepsilon(\mathbf{p-k-q}) (t-t^\prime)} \theta(t-t^\prime).\label{FirstArbuz}
\end{align}
Here $\mathbf{p}$ is the external momentum, $\zeta$ takes the value 1 in the bosonic case and $3/2$ in fermionic case\footnote{This factor in the symmetry coefficient is related to the presence of summation over spin indices in the fermionic case.}.
Considering this diagram at zero external momentum and zero frequency we get
\begin{align}
&\raisebox{-1.4em}{\begin{tikzpicture}
\begin{feynman}
\vertex (a1) ;
\vertex[right=0.9cm of a1] (a2);
\vertex[right=0.9cm of a2] (a3);
\vertex[right=0.9cm of a3] (a4);
\vertex[right=0.9cm of a4] (a5);
\vertex[below=0.05cm of a5]
(a6);
\vertex[right=0.05cm of a6]
(a7);
\diagram* {
(a4) -- (a5), 
(a4) -- (a3),
(a4) -- [insertion=0.8,anti fermion={[size=1t]}10](a2),
(a2) -- [anti fermion,insertion=0.15](a1),
(a4) -- [out=-90, in=-90, looseness=1,insertion=0.2] (a2),
(a4) -- [out=90, in=90, looseness=1,insertion=0.2] (a2)
};
\end{feynman}
\end{tikzpicture}}
\Bigg|_{\mathbf{p}=0,~\omega=0}=
\int\limits_{-\infty}^\infty d(t-t^\prime)~\raisebox{-1.4em}{\begin{tikzpicture}
\begin{feynman}
\vertex (a1) ;
\vertex[right=0.9cm of a1] (a2);
\vertex[right=0.9cm of a2] (a3);
\vertex[right=0.9cm of a3] (a4);
\vertex[right=0.9cm of a4] (a5);
\vertex[below=0.05cm of a5]
(a6);
\vertex[right=0.05cm of a6]
(a7);
\diagram* {
(a4) -- (a5), 
(a4) -- (a3),
(a4) -- [insertion=0.8,anti fermion={[size=1t]}10](a2),
(a2) -- [anti fermion,insertion=0.15](a1),
(a4) -- [out=-90, in=-90, looseness=1,insertion=0.2] (a2),
(a4) -- [out=90, in=90, looseness=1,insertion=0.2] (a2)
};
\end{feynman}
\end{tikzpicture}}\Bigg|_{\mathbf{p}=0}=\notag\\
&=\frac{\mp \zeta g^2}{(2\pi)^6}\int d\mathbf{k}d\mathbf{q}\bigg(\pi\delta\big(\varepsilon(\mathbf{k+q})-\varepsilon(\mathbf{k})-\varepsilon(\mathbf{q})\big)+\frac{i}{\varepsilon(\mathbf{k+q})-\varepsilon(\mathbf{k})-\varepsilon(\mathbf{q})}\bigg)\times\notag\\
&\times\big(1\pm 2n(\mathbf{k})\big)\big(1\pm 2n(\mathbf{q})\big).
\end{align}
The imaginary part of this integral contains the singularity on the surface given by the equation $\varepsilon(\mathbf{k+q})-\varepsilon(\mathbf{k})-\varepsilon(\mathbf{q})=0$. 
Such anomalous behavior at large times is called a pinch singularity. It can be shown by a straightforward calculation \cite{ELSM} that such singularities of order $g^2$ are contained in the diagrams\footnote{With external lines similar to the diagram (\ref{FirstArbuz}).}
\begin{equation}
\begin{tikzpicture}
\begin{feynman}
\vertex (a1) ;
\vertex[right=0.9cm of a1] (a2);
\vertex[right=0.9cm of a2] (a3);
\vertex[right=0.9cm of a3] (a4);
\vertex[right=0.9cm of a4] (a5);
\vertex[below=0.05cm of a5]
(a6);
\vertex[right=0.05cm of a6]
(a7) {\(,\)};
\diagram* {
(a4) -- (a5), 
(a4) -- (a3),
(a4) -- [insertion=0.8,anti fermion={[size=1t]}10](a2),
(a2) -- [anti fermion,insertion=0.15](a1),
(a4) -- [out=-90, in=-90, looseness=1,insertion=0.2] (a2),
(a4) -- [out=90, in=90, looseness=1,insertion=0.2] (a2)
};
\end{feynman}
\end{tikzpicture} \qquad
\begin{tikzpicture}
\begin{feynman}
\vertex (a1) ;
\vertex[right=0.9cm of a1] (a2);
\vertex[right=0.9cm of a2] (a3);
\vertex[right=0.9cm of a3] (a4);
\vertex[right=0.9cm of a4] (a5);
\vertex[below=0.05cm of a5]
(a6);
\vertex[right=0.05cm of a6]
(a7) {\(,\)};
\diagram* {
(a4) -- (a5), 
(a4) -- (a3),
(a4) -- [insertion=0.2](a2),
(a2) -- [anti fermion,insertion=0.15](a1),
(a4) -- [out=-90, in=-90, looseness=1,insertion=0.8] (a2),
(a4) -- [out=90, in=90, looseness=1,insertion=0.2,anti fermion={[size=1t]}10] (a2)
};
\end{feynman}
\end{tikzpicture}\qquad
\begin{tikzpicture}
\begin{feynman}
\vertex (a1) ;
\vertex[right=0.9cm of a1] (a2);
\vertex[right=0.9cm of a2] (a3);
\vertex[right=0.9cm of a3] (a4);
\vertex[right=0.9cm of a4] (a5);
\vertex[below=0.05cm of a5]
(a6);
\vertex[right=0.05cm of a6]
(a7) {\(.\)};
\diagram* {
(a4) -- (a5), 
(a4) -- (a3),
(a4) -- [insertion=0.8,anti fermion={[size=1t]}10](a2),
(a2) -- [anti fermion,insertion=0.15](a1),
(a4) -- [out=-90, in=-90, looseness=1,insertion=0.2,anti fermion={[size=1t]}10] (a2),
(a4) -- [out=90, in=90, looseness=1,insertion=0.2,anti fermion={[size=1t]}10] (a2)
};
\end{feynman}
\end{tikzpicture}\notag
\end{equation}
We denote the sum of the above two-loop diagrams in the momentum~-- frequency representation as $I(\mathbf{p},~\omega).$
It can also be shown that the sums over the “ear” type diagrams $$\feynmandiagram [horizontal=a to b, layered layout] {
  a -- b [dot] -- [out=135, in=45, loop, min distance=3cm] b -- c,
};$$
are free from pinch singularities. We do not consider these diagrams, since they are only additions to the chemical potential

In order to
deal with the pinch singularities we introduce the regulator  $\gamma$ in the propagators \cite{osnovnaya, ELSM}
\begin{equation}
\label{reg}
e^{-i\varepsilon(t-t')}\rightarrow e^{-i\varepsilon(t-t')-\gamma|t-t'|}.
\end{equation}
The introduction of this type of a regulator corresponds to the addition of a weak attenuation to the system.
It is convenient to choose the attenuation parameter as a function of momentum $\gamma=\gamma(\mathbf{p})$. We choose the attenuation parameter to be of the form $\gamma(\mathbf{p})=\alpha\mathbf{p}^2/2m,$ where the dimensionless parameter $\alpha\rightarrow 0$ when the regularization is removed.
Physically, this choice of $\gamma(\mathbf{p})$ form reflects the fact that the renormalized parameter $\alpha$ is analogous to the hydrodynamical kinetic Onsager coefficient 
\cite{Landau, Vasiljev2}. For example, the widely known A model of critical dynamics has the same form of exponential time decay \cite{Vasiljev2,5loops,5loops2,tauber,tauber2}. Moreover, in A model the role of $\alpha$ is played by the Onsager coefficient\footnote{Up to the factor $1/2m$.}. This is the motivation for our choice of the $\gamma(\mathbf{p})$ form. In the Keldysh~-- Schwinger formalism, the analogue of hydrodynamic equations are the Schwinger equations (see \cite{Fmodel}). 
After introducing the regularization, the free part of the action can be written in the form \cite{N}
\begin{equation}
\label{DeySvobReg}
S_0=-i\eta^+\big( 2\gamma(1\pm 2n)\big)\eta-i\xi^+\big(\partial_{t}+i\varepsilon-\gamma\big)\eta-i\eta^+\big(\partial_{t}+i\varepsilon+\gamma\big)\xi.
\end{equation}

A natural tool for the analysis of the attenuation is the Dyson equation
\cite{Vasiljev1,Vasiljev2}
\begin{equation}
\label{Dyson}
D^{-1}=K-\Sigma,
\end{equation}
where $D$ is the matrix of the Green's functions, $K$ is the matrix of the action quadratic form, $\Sigma$ is the self energy. The idea is to find a loop contribution to the regulator\footnote{We call this procedure “dressing”.} (in the limit $\alpha \to 0$), using the Dyson equations (\ref{Dyson}). With our choose of  $\gamma(\mathbf{p})$ this contributions need to be real and proportional to $\mathbf{p}^2$. If such a contribution exists, then the full Green's functions have an exponential in time decay (\ref{reg}) even after removing the regularization.
Notice that the self-energy may contain corrections of higher orders in $\mathbf{p}^2$, which give extra attenuation factors. However, since we are interested in the large-scale behavior in the regime of small $\mathbf{p}$ and $\omega$ (the hydrodynamic limit), the higher corrections can be neglected.

For carrying out the “dressing” procedure it is convenient to consider the matrix element
\begin{equation}
\label{Dyson1}
D^{-1}_{\eta^+\xi}=i\omega+i\varepsilon+\gamma-\Sigma_{\eta^+\xi},
\end{equation}
where $D^{-1}_{\eta^+\xi},~\Sigma_{\eta^+\xi}$ are the matrix elements of $D^{-1}$ and $\Sigma$, related to the fields $\eta^+$ and $\xi$.
For the reasons given above, it makes sense to consider the expression for $\Sigma_{\eta^+\xi}$ in the hydrodynamic limit, approximating its self energy by its Taylor series in the vicinity of $\omega = 0,~\mathbf{p}=0$. To analyze the effects of dissipation, it is sufficient to keep only the first nontrivial terms of the expansion proportional to $\omega$ and $\mathbf{p}^2$.
After passing the limit $\alpha\rightarrow 0$ the equation (\ref{Dyson1}) in hydrodynamic limit may be written in the form
\begin{equation}
\label{DysonRen}
D^{-1}_{\eta^+\xi}=(1+a_1+ia_2)i\omega+(i+ib_1+b_2)\frac{\mathbf{p}^2}{2m}-i\mu(c_1+ic_2)
\end{equation}
in $g^2$ order of perturbation theory. Here $a_i,b_i,c_i,~i\in\{1,2\}$ are some functions of arguments $T$ and $\mu$. It follows from the $\Sigma_{\eta^+\xi}$ loop contributions (see Appendix A). Since the loop corrections we are interested in are proportional to $g^2$, the contributions of $a_1$ and $b_1$ can be neglected in the equation (\ref{DysonRen}).
After that, we define the “dressed” attenuation parameter in $g^2$ order as
\begin{equation}
\label{abc}
\tilde\alpha = \text{Re}\bigg(\frac{i+ib_1+b_2}{1+a_1+ia_2}\bigg) \approx b_2+a_2.
\end{equation}
Note that the contributions responsible for the emergence of the imaginary part of the chemical potential are possible. However, chemical potential can not have the imaginary part according to Neumann equation. The resolution of the apparent paradox lies in the fact that the Dyson equation is, in fact, a self-consistency equation. Thus, one needs to add some terms in the Dyson equation that ensure that the chemical potential is real.

In accordance with the logic of the hydrodynamic description let us expand $I(\mathbf{p},~\omega)$ in a power series around $\mathbf{p}=0,~\omega=0$:
\begin{equation}
\label{Iorig}
I(\mathbf{p},~\omega)\approx I(\mathbf{p}=0,~\omega=0)+\omega\cdot\frac{\partial I(\mathbf{p}=0,~\omega)}{\partial\omega}\Bigg|_{\omega=0}+\frac{p_i p_j}{2!}\cdot \frac{\partial^2 I(\mathbf{p},~\omega=0)}{\partial p_i \partial p_j}\Bigg|_{\mathbf{p}=0},
\end{equation}
where $p_i$ is the $i$-th component of $\mathbf{p}$. A sum over the repeated indices is implied. There are no terms in the expansion linear in the momentum components in the cause of homogeneity of the system.
It is also possible to significantly simplify the further analysis by
substituting $\mu=0$ into the denominators of the integrands of all the terms in the expansion (\ref{Iorig}). This is possible if we switch from considering the hydrodynamic variable $\omega$ to the more natural hydrodynamic variable $\omega-\mu$.

To pass to the zero frequency in the momentum-time representation it is sufficient to integrate the corresponding diagrams over the time difference in infinite limits.
After this procedure, the sum of the diagrams at $\omega=0$ is given by (see. \mbox{Appendix B)}
\begin{align}
I(\mathbf{p},\omega=0)=\frac{\mp 2\zeta g^2}{(2\pi)^6}\int d\mathbf{k}d\mathbf{q}~\frac{2n(\mathbf{k})n(\mathbf{q})\mp 2n(\mathbf{p-k-q})-4n(\mathbf{k})n(\mathbf{p-k-q})}{i\big(\tilde\varepsilon(\mathbf{p-k-q})-\tilde\varepsilon(\mathbf{k})-\tilde\varepsilon(\mathbf{q})\big)-\gamma(\mathbf{k})-\gamma(\mathbf{q})-\gamma(\mathbf{p-k-q})}\label{DiagP}.
\end{align}
Here $\tilde\varepsilon(\mathbf{p})=\mathbf{p}^2/2m$. Analogously, the derivative of $I(\mathbf{p},\omega)$ with respect to frequency can be written in the form
\begin{align}
\frac{\partial I(\mathbf{p},\omega)}{\partial\omega}\Bigg|_{\omega=0}=\frac{\mp 2i\zeta g^2}{(2\pi)^6}\int d\mathbf{k} d\mathbf{q}~\frac{2n(\mathbf{k})n(\mathbf{q})\mp 2n(\mathbf{p-k-q})-4n(\mathbf{k})n(\mathbf{p-k-q})}{\big[i\big(\tilde\varepsilon(\mathbf{p-k-q})-\tilde\varepsilon(\mathbf{k})-\tilde\varepsilon(\mathbf{q})\big)-\gamma(\mathbf{k})-\gamma(\mathbf{q})-\gamma(\mathbf{p-k-q})\big]^2} \label{DiagOmega}.
\end{align}
Here the factor $i$ appears as a result of the transition from the momentum~-- frequency representation to the momentum~-- time representation.
The term  $I(\mathbf{p}=0,~\omega=0)$ is not interesting for us because in the Dyson equation this term is only an addition to the chemical potential.

Thus, taking the necessary derivatives,
the contributions from (\ref{Iorig}) give that the “dressed” attenuation parameter can be written as\footnote{We give these expressions up to contributions which goes to 0 when $\alpha\rightarrow 0$.}:
\begin{align}
&\frac{\partial I(\mathbf{p}=0,~\omega)}{\partial\omega}\Bigg|_{\omega=0}=\frac{\mp 2i\zeta g^2m^2}{(2\pi)^6}\int d\mathbf{k} d\mathbf{q}~\frac{f(\mathbf{k}, \mathbf{q})}{\big[i(\mathbf{k}\cdot\mathbf{q})-\alpha(\mathbf{k}^2+\mathbf{k\cdot q}+\mathbf{q}^2)\big]^2},\label{IntLyn}\\
&\frac{\partial^2 I(\mathbf{p},~\omega=0)}{\partial p_i \partial p_j}\Bigg|_{\mathbf{p}=0}=\delta_{ij}\frac{\mp2 \zeta g^2m}{(2\pi)^6 }\int d\mathbf{k} d\mathbf{q}\Bigg(-\frac{ if(\mathbf{k},\mathbf{q})}{\big[i(\mathbf{k\cdot q})-\alpha(\mathbf{k}^2+\mathbf{k\cdot q}+\mathbf{q}^2)\big]^2}+ \notag\\
&+\frac{2}{3}\frac{ f(\mathbf{k},\mathbf{q})\cdot(\mathbf{k+q})^2}{\big[i(\mathbf{k\cdot q})-\alpha(\mathbf{k}^2+\mathbf{k\cdot q}+\mathbf{q}^2)\big]^3}+\frac{f_1(\mathbf{p},\mathbf{k})}{i(\mathbf{k\cdot q})-\alpha(\mathbf{k}^2+\mathbf{k\cdot q}+\mathbf{q}^2)}+\notag\\&+
\frac{2}{3}\frac{ if_2(\mathbf{k},\mathbf{q})\cdot(\mathbf{k+q})^2}{\big[i(\mathbf{k\cdot q})-\alpha(\mathbf{k}^2+\mathbf{k\cdot q}+\mathbf{q}^2)\big]^2}\Bigg)\label{IntKvadr}.
\end{align}
Here
$$f(\mathbf{p},\mathbf{k},\mathbf{q})=2n(\mathbf{k})n(\mathbf{q})\mp 2n(\mathbf{p-k-q})-4n(\mathbf{k})n(\mathbf{p-k-q}),$$ $$f(\mathbf{k},\mathbf{q})=f(\mathbf{p},\mathbf{k},\mathbf{q})\big|_{\mathbf{p}=0},$$ $$\nonumber \partial_{p_i}  \partial_{p_j}        f(\mathbf{p},\mathbf{k},\mathbf{q})\big|_{\mathbf{p}=0}\equiv f_1(\mathbf{k},\mathbf{q})\cdot \delta_{ij},\quad\partial_{p_j}        f(\mathbf{p},\mathbf{k},\mathbf{q})\big|_{\mathbf{p}=0}\equiv(\mathbf{k}_j+\mathbf{q}_j)\cdot f_2(\mathbf{k},\mathbf{q}),$$ $\delta_{ij}$ is the Kronecker delta, $\mathbf{k}_i$ and $\mathbf{q}_i$ are the $i$-th components of the vectors $\mathbf{k}$ and $\mathbf{q}$. For homogeneity, 
all terms $(\mathbf{k}_i+\mathbf{q}_i)\cdot(\mathbf{k}_j+\mathbf{q}_j)$ in the integrands were replaced by $\delta_{ij}/3\cdot(\mathbf{k+q})^2$. 
Note that the real part of (\ref{IntKvadr}) and (\ref{IntLyn}) arises due to the introduction of the regularization (\ref{reg}). The corresponding expressions for $ \alpha = 0 $ do not contain the real part.

\begin{figure}[h]  
\vspace{4ex} \centering \subfigure[]{
\includegraphics[width=0.455\linewidth]{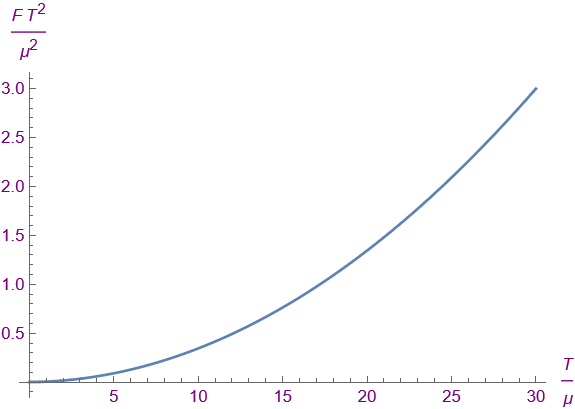} \label{fig:fermions1}}  
\hspace{4ex}
\subfigure[]{
\includegraphics[width=0.455\linewidth]{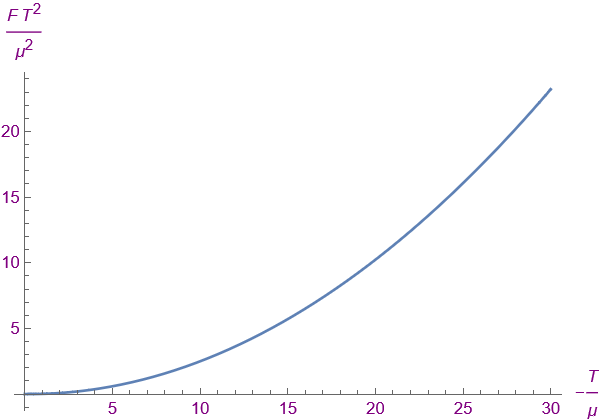}
\label{fig:fermions2} }
\hspace{4ex}
\subfigure[]{ \includegraphics[width=0.455\linewidth]{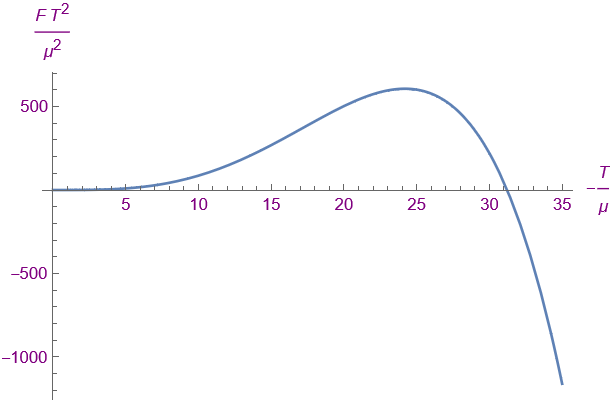} \label{fig:bose} }  
\caption{The function $F(T/\mu)$ in case of \subref{fig:fermions1} fermions with positive $\mu$; \subref{fig:fermions2} fermions with negative $\mu$; \subref{fig:bose} bosons with negative $\mu$.} \label{fig:bosons}
\end{figure}

Following the logic above, it is necessary to find a term of the asymptotics of the real part of the integrals (\ref{IntLyn}-\ref{IntKvadr}) that does not disappear when the regularization is removed. According to (\ref{abc}), this real part contribute to the $\gamma$ renormalization. This asymptotic analysis is rather complicated due to the extremely nonuniform behavior of  the integrals (\ref{IntLyn}-\ref{IntKvadr}) with respect to the $\alpha$ parameter. At the moment, there is no general theory for such type of integrals. Integrals of a similar type were considered in the articles \cite{Ershov1, Ershov2,Ilin}. 
It can be shown (see Appendix A) that in the limit $\alpha\rightarrow 0$ the real parts of (\ref{IntKvadr}) and (\ref{IntLyn}) behave as\footnote{That is, they do not go to 0 when $\alpha\rightarrow 0$.} $\mathcal{O}(1)$. The explicit form of the “dressed” attenuation parameter can be written as 
\begin{equation}
\tilde\alpha=g^2 m^2 T^{2} F\big(T/\mu\big),
\end{equation}
where $F\big(T/\mu\big)$ is the dimensionless function. Its numerical values for the bosonic and fermionic cases are presented in the Figures \ref{fig:fermions1}-\ref{fig:bose} with the extra factor $T^2/\mu^2$. 
It can be seen that in the bosonic case there is a region of the parameter $T/\mu$ where $\tilde\alpha$ becomes negative. These values of $\tilde{\alpha}$ are not possible due to the instability of the system. The description of this region is apparently impossible due to the presence of a phase transition ($\mu=0$ at $T\neq 0$).
The case of bosons with a positive chemical potential is not considered for standard reasons.

\section{Conclusion}
In this paper we presented a microscopic justification of dissipation within the model of the equilibrium non-relativistic Fermi or Bose many-body system with the weak local interactions. The results of works \cite{osnovnaya,ELSM} were generalized to the case of an arbitrary equilibrium state above the critical point. It was shown that the time decay of 2-point Green's functions arises from the loop contributions which contain pinch singularities. The natural way to show this is to introduce a regularization that provides exponential decay of the propagators and to “dress” the regularization parameter using the Dyson equation.
The form of this attenuation is chosen similar to well-known semi-phenomenological hydrodynamic models (for example, model A of critical dynamics). In accordance with this, the “dressed” regularization parameter $\tilde\alpha$, which we called the attenuation parameter, can be considered a microscopic analogue of the Onsager kinetic coefficient. It is shown that in this model this parameter is proportional to $\mathcal{O}(g^2)$, i.e. these effects appear in the two-loop approximation. This explains why the dissipation-type effects have not been discovered earlier in the framework of one-loop approximations (such as the Hartree or Hartree~-- Fock \cite{Mattuk}). The attenuation parameter was calculated as a function of temperature and chemical potential in $g^2$ order. When calculating $\tilde\alpha$, the main difficulty was the asymptotic analysis of a certain class of integrals that depend extremely nonuniformly on the regularization parameter. Note that the construction of a general theory for this types of integrals can be an interesting problem in the framework of asymptotic analysis.

The technique of time-dependent Green's functions at a finite temperature provides ample opportunities for calculating damping characteristics in the case of more realistic interaction potentials in weakly coupled Fermi and Bose systems. It is possible to use the described formalism in other models of statistical mechanics and condensed matter physics. For example, it seems prospective to consider from this point of view the dissipative regime of a system of two-dimensional massless relativistic fermions, which is a low-energy approximation for a system of conduction electrons in graphene \cite{Graphene19,Novoselov05}. 
\\ \\
\textbf{Acknowledgements}.
\\ \\
Viacheslav Krivorol would like to thank Ivan Vybornyi, Taisiia Zaitseva and Andrei Ovsiannikov for their help in preparing the arXiv version of this text.
\\ \\
This research was supported by the Foundation for the Advancement of Theoretical Physics and Mathematics “BASIS (Grant No. 19-1-1-35-1)”.

\newpage
\section*{Appendix A. Asymptotic analysis of the limit $\alpha\rightarrow 0$}
Let us simplify the integrals (\ref{IntLyn}-\ref{IntKvadr}). For example, we consider the simplest contribution to (\ref{IntKvadr}),
\begin{equation}
\label{I1}
I_1=\frac{\mp 2\zeta g^2m}{(2\pi)^6 }\int\frac{        f_1(\mathbf{k},\mathbf{q})}{i(\mathbf{k\cdot q})-\alpha(\mathbf{k}^2+\mathbf{k\cdot q}+\mathbf{q}^2)}d\mathbf{k} d\mathbf{q}.
\end{equation}
Another contributions to (\ref{IntLyn}) and (\ref{IntKvadr})
\begin{align}
&I_2= \frac{\pm 2\zeta g^2m}{(2\pi)^6 }\int d\mathbf{k} d\mathbf{q}\frac{i f(\mathbf{k},\mathbf{q})-\frac{2}{3}(\mathbf{k+q})^2\cdot        if_1(\mathbf{k},\mathbf{q})}{\big[i(\mathbf{k\cdot q})-\alpha(\mathbf{k}^2+\mathbf{k\cdot q}+\mathbf{q}^2)\big]^2},\\
&I_3=\frac{\mp 4\zeta g^2m}{3(2\pi)^6 }\int d\mathbf{k} d\mathbf{q}\frac{ f(\mathbf{k},\mathbf{q})\cdot(\mathbf{k+q})^2}{\big[i(\mathbf{k\cdot q})-\alpha(\mathbf{k}^2+\mathbf{k\cdot q}+\mathbf{q}^2)\big]^3},\\
&I_4=\frac{\mp 2\zeta g^2m^2}{(2\pi)^6}\int d\mathbf{k} d\mathbf{q}~\frac{f(\mathbf{k}, \mathbf{q})}{\big[i(\mathbf{k}\cdot\mathbf{q})-\alpha(\mathbf{k}^2+\mathbf{k\cdot q}+\mathbf{q}^2)\big]^2}
\end{align}
are simplified in the same way.
Consider the real part of (\ref{I1}):
\begin{equation}
\Re (I_1)=\frac{\pm 2\zeta g^2m}{(2\pi)^6 }\int\frac{\alpha (\mathbf{k}^2+\mathbf{k\cdot q}+\mathbf{q}^2)~f_1(\mathbf{x},\mathbf{y})}{(\mathbf{k\cdot q})^2+\alpha^2(\mathbf{k}^2+\mathbf{k\cdot q}+\mathbf{q}^2)^2}d\mathbf{k} d\mathbf{q}.
\end{equation}
Do a linear change of the variables $$\mathbf{k}=\frac{1}{\sqrt{2}}\mathbf{(x+y)},\quad\mathbf{q}=\frac{1}{\sqrt{2}}\mathbf{(x-y)}.$$ There are no scalar products of different vectors in these variables,
\begin{equation}
\label{Ipromejut}
\Re(I_1)=\frac{\pm 4\zeta g^2m}{(2\pi)^6}\int d\mathbf{x}d\mathbf{y}\frac{ \alpha(3\mathbf{x}^2+\mathbf{y}^2)f_1(\mathbf{x},\mathbf{y})}{(\mathbf{x}^2-\mathbf{y}^2)^2+\alpha^2(3\mathbf{x}^2+\mathbf{y}^2)}.
\end{equation}
Hereinafter we accept the convention that after a coordinate transformation the functions will be denoted by the same letter, fixing the changes by replacing the old arguments with the new ones. For example, here $f_1(\mathbf{k},\mathbf{q})\rightarrow f_1(\mathbf{x},\mathbf{y}).$

Use spherical coordinates. The integrand in (\ref{Ipromejut}) depends on $x=|\mathbf{x}|,~y=|\mathbf{y}|$ and angle $\theta$ between $\mathbf{x}$ and $\mathbf{y}$. Integrals over other variables can be calculated trivially,
\begin{equation}
\label{Ipromej2}
\Re(I_1)=\frac{\pm \zeta g^2m}{2\pi^4}\int\limits_0^\pi d\theta \sin{\theta}\int\limits_0^\infty\int\limits_0^\infty dx dy~ \frac{\alpha x^2y^2(3x^2+y^2)f_1(x,y,\theta)}{(x^2-y^2)^2+\alpha^2(3x^2+y^2)^2}.
\end{equation}
Change the variables $\{x^2=s,~y^2=t,~\cos\theta=c\}$ in (\ref{Ipromej2})
\begin{equation}
\Re(I_1)=\frac{\pm 8\zeta g^2 m}{89\pi^4}\int\limits_{-1}^1dc\int\limits_0^\infty\int\limits_0^\infty ds dt~ \frac{\alpha\sqrt{st}(3s+t) f_1(s,t,c)}{(s-t)^2+\alpha^2(3s+t)^2}.
\end{equation}
Integrating by parts over $s$ and passing the limit $\alpha\rightarrow 0$ we get
\begin{equation}
\Re(I_1)=\frac{\mp 4\zeta g^2 m}{89\pi^3}\int\limits_{-1}^1dc\int\limits_0^\infty\int\limits_0^\infty ds dt~\text{sign}(s-t)\sqrt{t}\partial_s\Big(\sqrt{s} f_1(s,t)\Big).
\end{equation}
Here $\text{sign}(s-t)$ is the sign function.
The boundary terms are equal to 0 on limits $s=0$ and $s=\infty$\footnote{The function $f_1$ as function of $s$ decays exponentially at the infinity.}.
Using one more integration by parts and translating the derivative with respect to $s$ to the sign function, the integral can be rewritten as
\begin{equation}
\label{I1st}
\Re(I_1)=\frac{\pm 8\zeta g^2 m}{89\pi^3}\int\limits_{-1}^1dc\int\limits_0^\infty\int\limits_0^\infty ds dt~\delta(s-t)\sqrt{st}~f_1(s,t).
\end{equation}
The other integrals are simplified in the similar way. It is only necessary to increase the number of integrations by parts over such variables so that nontrivial boundary terms do not arise. Making these calculations we get\footnote{Note that according to (\ref{abc}) it is the imaginary part of $I_4$ that contributes to the damping.}:
\begin{align}
\label{I2st}
&\Re(I_2)=\frac{\mp \zeta g^2 m}{12\pi^3}\int\limits_{-1}^1dc\int\limits_0^\infty\int\limits_0^\infty ds dt~\delta(s-t)\sqrt{t}~\partial_s\Big(\sqrt{s}\big(3f(s,t)-4sf_2(s,t)\big)\Big),\\
\label{I3st}
&\Re(I_3)=\frac{\pm \zeta g^2 m}{3\pi^3}\int\limits_{-1}^1dc\int\limits_0^\infty\int\limits_0^\infty ds dt~\delta(s-t)\sqrt{t}~\partial_t\partial_s\Big(s^{3/2}\sqrt{t}f(s,t)\Big),\\
&\Im(I_4)=\frac{\pm \zeta g^2 m^2}{4\pi^3}\int\limits_{-1}^1dc\int\limits_0^\infty\int\limits_0^\infty ds dt~\delta(s-t)\partial_s\partial_t\Big(\sqrt{st}f(s,t)\Big).
\end{align}
Integrals over $c$ can be calculated analytically using computer algebra systems\footnote{Mathematica, Maple and others.}.
Then, in (\ref{I1st}-\ref{I3st}), if $\mu>0$, it is necessary to calculate the integration over the variable $s$ using the $\delta$-function. If $\mu<0$, it is necessary to do it over the variable $ t$. This allows us to avoid the emergence of square roots from negative expressions. After performing these integrations the problem is reduced to the calculation of one-dimensional integrals. It is convenient to write the sum of the remaining expressions as the product of a dimensionless function and the dimensional prefactor:
\begin{equation}
m\cdot\Re(I_1+I_2+I_3)+\Im(I_4)=g^2 m^2 T^{2} F\big(T/\mu\big).
\end{equation}
Here the integrals are nondimensionalized by rescaling of the variables $\{s,t\}\rightarrow\{m s/\beta, m t/\beta\}$.
After that the function $F(T/\mu)$ can be calculated numerically as a one-dimensional integral for each value of the parameter $T/\mu$. Its numerical values are given in the main text.

\section*{Appendix B. First coefficients of Feynman diagram expansions around $\omega=0$.}

In this section, the coefficients of the expansion in frequency in the zeroth and linear order of the Feynman diagrams that arise in the work are given. Linear coefficients are denoted here as $\{\ \cdot ~ \}_\omega$. The integration variable $t-t^\prime$ is denoted by $\tau$.
\begin{align}
&\raisebox{-1.4em}{\begin{tikzpicture}
\begin{feynman}
\vertex (a1) ;
\vertex[right=0.9cm of a1] (a2);
\vertex[right=0.9cm of a2] (a3);
\vertex[right=0.9cm of a3] (a4);
\vertex[right=0.9cm of a4] (a5);
\vertex[below=0.05cm of a5]
(a6);
\vertex[right=0.05cm of a6]
(a7);
\diagram* {
(a4) -- (a5), 
(a4) -- (a3),
(a4) -- [insertion=0.8,anti fermion={[size=1t]}10](a2),
(a2) -- [anti fermion,insertion=0.15](a1),
(a4) -- [out=-90, in=-90, looseness=1,insertion=0.2] (a2),
(a4) -- [out=90, in=90, looseness=1,insertion=0.2] (a2)
};
\end{feynman}
\end{tikzpicture}}\Bigg|_{\omega=0}=\frac{\mp \zeta g^2}{(2\pi)^6}\int\limits_0^\infty d\tau\int d\mathbf{k}d\mathbf{q} \big(1\pm 2n(\mathbf{k})\big)\big(1\pm 2n(\mathbf{q})\big)\times~~~~~~~ \notag \\
&\times\exp{\Big(i\tau\big(\varepsilon(\mathbf{p-k-q})-\varepsilon(\mathbf{k})-\varepsilon(\mathbf{q})\big)-|\tau|\big(\gamma(\mathbf{k})+\gamma(\mathbf{q})+\gamma(\mathbf{p-k-q})\big)\Big)}=\notag\\
&=\frac{\mp \zeta g^2}{(2\pi)^6}\int d\mathbf{k}d\mathbf{q}~\frac{\big(1\pm 2n(\mathbf{k})\big)\big(1\pm 2n(\mathbf{q})\big)}{i\big(\varepsilon(\mathbf{p-k+q})-\varepsilon(\mathbf{k})-\varepsilon(\mathbf{q})\big)-\gamma(\mathbf{k})-\gamma(\mathbf{q})-\gamma(\mathbf{p-k-q})},\\
&\raisebox{-1.4em}{\begin{tikzpicture}
\begin{feynman}
\vertex (a1) ;
\vertex[right=0.9cm of a1] (a2);
\vertex[right=0.9cm of a2] (a3);
\vertex[right=0.9cm of a3] (a4);
\vertex[right=0.9cm of a4] (a5);
\vertex[below=0.05cm of a5]
(a6);
\vertex[right=0.05cm of a6]
(a7) {\(\)};
\diagram* {
(a4) -- (a5), 
(a4) -- (a3),
(a4) -- [insertion=0.2](a2),
(a2) -- [anti fermion,insertion=0.15](a1),
(a4) -- [out=-90, in=-90, looseness=1,insertion=0.8] (a2),
(a4) -- [out=90, in=90, looseness=1,insertion=0.2,anti fermion={[size=1t]}10] (a2)
};
\end{feynman}
\end{tikzpicture}}\Bigg|_{\omega=0}=\frac{\pm 2\zeta g^2}{(2\pi)^6}\int\limits_0^\infty d\tau\int d\mathbf{k}d\mathbf{q} \big(1\pm 2n(\mathbf{k})\big)\big(1\pm 2n(\mathbf{p-k-q})\big)\times~~~~~~~ \notag \\
&\times\exp{\Big(i\tau\big(\varepsilon(\mathbf{p-k-q})-\varepsilon(\mathbf{k})-\varepsilon(\mathbf{q})\big)-|\tau|\big(\gamma(\mathbf{k})+\gamma(\mathbf{q})+\gamma(\mathbf{p-k-q})\big)\Big)}=\notag\\
&=\frac{\pm 2\zeta g^2}{(2\pi)^6}\int d\mathbf{k}d\mathbf{q}~\frac{\big(1\pm 2n(\mathbf{k})\big)\big(1\pm 2n(\mathbf{p-k-q})\big)}{i\big(\varepsilon(\mathbf{p-k+q})-\varepsilon(\mathbf{k})-\varepsilon(\mathbf{q})\big)-\gamma(\mathbf{k})-\gamma(\mathbf{q})-\gamma(\mathbf{p-k-q})},\\
&\raisebox{-1.4em}{\begin{tikzpicture}
\begin{feynman}
\vertex (a1) ;
\vertex[right=0.9cm of a1] (a2);
\vertex[right=0.9cm of a2] (a3);
\vertex[right=0.9cm of a3] (a4);
\vertex[right=0.9cm of a4] (a5);
\vertex[below=0.05cm of a5]
(a6);
\vertex[right=0.05cm of a6]
(a7);
\diagram* {
(a4) -- (a5), 
(a4) -- (a3),
(a4) -- [insertion=0.8,anti fermion={[size=1t]}10](a2),
(a2) -- [anti fermion,insertion=0.15](a1),
(a4) -- [out=-90, in=-90, looseness=1,insertion=0.2,anti fermion={[size=1t]}10] (a2),
(a4) -- [out=90, in=90, looseness=1,insertion=0.2,anti fermion={[size=1t]}10] (a2)
};
\end{feynman}
\end{tikzpicture}}\Bigg|_{\omega=0}=\frac{\mp \zeta g^2}{(2\pi)^6}\int\limits_0^\infty d\tau\int d\mathbf{k}d\mathbf{q} \exp{\Big(i\tau\big(\varepsilon(\mathbf{p-k-q})-\varepsilon(\mathbf{k})-\varepsilon(\mathbf{q})\big)\Big)}\times \notag \\
&\times\exp{\Big(-|\tau|\big(\gamma(\mathbf{k})+\gamma(\mathbf{q})+\gamma(\mathbf{p-k-q})\big)\Big)}=\notag\\
&=\frac{\mp \zeta g^2}{(2\pi)^6}\int d\mathbf{k}d\mathbf{q}~\frac{1}{i\big(\varepsilon(\mathbf{p-k+q})-\varepsilon(\mathbf{k})-\varepsilon(\mathbf{q})\big)-\gamma(\mathbf{k})-\gamma(\mathbf{q})-\gamma(\mathbf{p-k-q})}.\\
&\raisebox{-1.4em}{\begin{tikzpicture}
\begin{feynman}
\vertex (a1) {\(\Bigg\{\)};
\vertex[right=1.2 cm of a1] (a2);
\vertex[right=0.9cm of a2] (a3);
\vertex[right=0.9cm of a3] (a4);
\vertex[right=0.9cm of a4] (a5);
\vertex[below=0.05cm of a5]
(a6);
\vertex[right=0.05cm of a6]
(a7) {\(\Bigg\}_{\omega}\)};
\diagram* {
(a4) -- (a5), 
(a4) -- (a3),
(a4) -- [insertion=0.8,anti fermion={[size=1t]}10](a2),
(a2) -- [anti fermion,insertion=0.15](a1),
(a4) -- [out=-90, in=-90, looseness=1,insertion=0.2] (a2),
(a4) -- [out=90, in=90, looseness=1,insertion=0.2] (a2)
};
\end{feynman}
\end{tikzpicture}}=\frac{\mp i\zeta g^2}{(2\pi)^6}\int\limits_0^\infty d\tau\int d\mathbf{k}d\mathbf{q}~\tau \big(1\pm 2n(\mathbf{k})\big)\big(1\pm 2n(\mathbf{q})\big)\times\nonumber\\
&\times\exp{\Big(i\tau\big(\varepsilon(\mathbf{k+q})-\varepsilon(\mathbf{k})-\varepsilon(\mathbf{q})\big)-|\tau|\big(\gamma(\mathbf{k})+\gamma(\mathbf{q})+\gamma(\mathbf{k+q})\big)\Big)}=\notag\\
&=\frac{\mp i\zeta g^2}{(2\pi)^6}\int d\mathbf{k}d\mathbf{q}~\frac{\big(1\pm 2n(\mathbf{k})\big)\big(1\pm 2n(\mathbf{q})\big)}{\Big[i\big(\varepsilon(\mathbf{p-k+q})-\varepsilon(\mathbf{k})-\varepsilon(\mathbf{q})\big)-\gamma(\mathbf{k})-\gamma(\mathbf{q})-\gamma(\mathbf{p-k-q})\Big]^2}.\\
&\raisebox{-1.4em}{\begin{tikzpicture}
\begin{feynman}
\vertex (a1)  {\(\Bigg\{\)};
\vertex[right=1.2cm of a1] (a2);
\vertex[right=0.9cm of a2] (a3);
\vertex[right=0.9cm of a3] (a4);
\vertex[right=0.9cm of a4] (a5);
\vertex[below=0.05cm of a5]
(a6);
\vertex[right=0.05cm of a6]
(a7) {\(\Bigg\}_{\omega}\)};
\diagram* {
(a4) -- (a5), 
(a4) -- (a3),
(a4) -- [insertion=0.2](a2),
(a2) -- [anti fermion,insertion=0.15](a1),
(a4) -- [out=-90, in=-90, looseness=1,insertion=0.8] (a2),
(a4) -- [out=90, in=90, looseness=1,insertion=0.2,anti fermion={[size=1t]}10] (a2)
};
\end{feynman}
\end{tikzpicture}}=\frac{\pm 2i\zeta g^2}{(2\pi)^6}\int\limits_0^\infty d\tau\int d\mathbf{k}d\mathbf{q}~\tau \big(1\pm 2n(\mathbf{k})\big)\big(1\pm 2n(\mathbf{k+q})\big)\times~~~~ \notag \\
&\times\exp{\Big(i\tau\big(\varepsilon(\mathbf{k+q})-\varepsilon(\mathbf{k})-\varepsilon(\mathbf{q})\big)-|\tau|\big(\gamma(\mathbf{k})+\gamma(\mathbf{q})+\gamma(\mathbf{k+q})\big)\Big)}=\notag\\
&=\frac{\pm 2i\zeta g^2}{(2\pi)^6}\int d\mathbf{k}d\mathbf{q}~\frac{\big(1\pm 2n(\mathbf{k})\big)\big(1\pm 2n(\mathbf{k+q})\big)}{\Big[i\big(\varepsilon(\mathbf{p-k+q})-\varepsilon(\mathbf{k})-\varepsilon(\mathbf{q})\big)-\gamma(\mathbf{k})-\gamma(\mathbf{q})-\gamma(\mathbf{p-k-q})\Big]^2}.\\
&\raisebox{-1.4em}{\begin{tikzpicture}
\begin{feynman}
\vertex (a1) {\(\Bigg\{\)};
\vertex[right=1.2cm of a1] (a2);
\vertex[right=0.9cm of a2] (a3);
\vertex[right=0.9cm of a3] (a4);
\vertex[right=0.9cm of a4] (a5);
\vertex[below=0.05cm of a5]
(a6);
\vertex[right=0.05cm of a6]
(a7) {\(\Bigg\}_{\omega}\)};
\diagram* {
(a4) -- (a5), 
(a4) -- (a3),
(a4) -- [insertion=0.8,anti fermion={[size=1t]}10](a2),
(a2) -- [anti fermion,insertion=0.15](a1),
(a4) -- [out=-90, in=-90, looseness=1,insertion=0.2,anti fermion={[size=1t]}10] (a2),
(a4) -- [out=90, in=90, looseness=1,insertion=0.2,anti fermion={[size=1t]}10] (a2)
};
\end{feynman}
\end{tikzpicture}}=\frac{\mp i\zeta g^2}{(2\pi)^6}\int\limits_0^\infty d\tau\int d\mathbf{k}d\mathbf{q}~\tau \times \nonumber \\
&\times\exp{\Big(i\tau\big(\varepsilon(\mathbf{k+q})-\varepsilon(\mathbf{k})-\varepsilon(\mathbf{q})\big)-|\tau|\big(\gamma(\mathbf{k})+\gamma(\mathbf{q})+\gamma(\mathbf{k+q})\big)\Big)}=\nonumber\\
&=\frac{\mp i\zeta g^2}{(2\pi)^6}\int d\mathbf{k}d\mathbf{q}~\frac{1}{\Big[i\big(\varepsilon(\mathbf{p-k+q})-\varepsilon(\mathbf{k})-\varepsilon(\mathbf{q})\big)-\gamma(\mathbf{k})-\gamma(\mathbf{q})-\gamma(\mathbf{p-k-q})\Big]^2}.
\end{align}

\newcommand{\eprint}[1]{\href{http://arxiv.org/abs/#1}{\texttt{#1}}}
\bibliographystyle{my3}
\bibliography{bibliography}

\begin{thebibliography}{10}
\newcommand{\enquote}[1]{``#1''}
\providecommand{\url}[1]{\texttt{#1}}
\providecommand{\urlprefix}{URL }
\expandafter\ifx\csname urlstyle\endcsname\relax
  \providecommand{\doi}[1]{doi:\discretionary{}{}{}#1}\else
  \providecommand{\doi}{doi:\discretionary{}{}{}\begingroup
  \urlstyle{rm}\Url}\fi
\providecommand{\eprint}[1]{\href{http://arxiv.org/abs/#1}{\texttt{#1}}}

\bibitem{zubarev1}
D.N. Zubarev, V.~Morozov, G.~Ropke, \enquote{Statistical Mechanics of
  Nonequilibrium Processes, Volume 1 (See 3527400834): Basic Concepts, Kinetic
  Theory}, Statistical Mechanics of Nonequilibrium Processes, Wiley, 1996.

\bibitem{zubarev2}
D.~Zubarev, V.~Morozov, G.~R{\"o}pke, \enquote{Statistical Mechanics of
  Nonequilibrium Processes, Statistical Mechanics of Nonequilibrium Processes.
  Volume 2: Relaxation and Hydrodynamic Processes}, Statistical Mechanics of
  Nonequilibrium Processes, Wiley.

\bibitem{snowmass}
Tomas Brauner, Sean Hartnoll, Pavel Kovtun, et~al., \enquote{Snowmass White
  Paper: Effective Field Theories for Condensed Matter Systems}, \emph{ArXiv
  preprint 2203.10110}.

\bibitem{HeavyIon}
Marcus Bluhm, et~al., \enquote{{Dynamics of critical fluctuations: Theory
  \textendash{} phenomenology \textendash{} heavy-ion collisions}},
  \href{http://dx.doi.org/10.1016/j.nuclphysa.2020.122016}{\emph{Nucl. Phys.
  A}}, \textbf{1003} (2020), 122016, \eprint{2001.08831}.

\bibitem{FEYNMAN1}
R.P Feynman, F.L Vernon, \enquote{The theory of a general quantum system
  interacting with a linear dissipative system},
  \href{http://dx.doi.org/https://doi.org/10.1016/0003-4916(63)90068-X}{\emph{Annals
  of Physics}}, \textbf{24} (1963), 118--173.

\bibitem{PATRIARCA}
Marco Patriarca, \enquote{Feynman–Vernon model of a moving thermal
  environment},
  \href{http://dx.doi.org/https://doi.org/10.1016/j.physe.2005.05.021}{\emph{Physica
  E: Low-dimensional Systems and Nanostructures}}, \textbf{29}: 1 (2005),
  243--250, frontiers of Quantum.

\bibitem{weiss}
Ulrich Weiss, \enquote{Quantum Dissipative Systems}, WORLD SCIENTIFIC, 3rd
  edn., 2008, \doi{10.1142/6738},
  \eprint{https://www.worldscientific.com/doi/pdf/10.1142/6738},
  \urlprefix\url{https://www.worldscientific.com/doi/abs/10.1142/6738}.

\bibitem{OpenQuantumIntr}
{\'A}.~Rivas, S.F. Huelga, \enquote{Open Quantum Systems: An Introduction},
  SpringerBriefs in Physics, Springer Berlin Heidelberg, 2011,
  \urlprefix\url{https://books.google.ru/books?id=FGCuYsIZAA0C}.

\bibitem{LindbladFirstPaper}
Goran Lindblad, \enquote{{On the Generators of Quantum Dynamical Semigroups}},
  \href{http://dx.doi.org/10.1007/BF01608499}{\emph{Commun. Math. Phys.}},
  \textbf{48} (1976), 119.

\bibitem{Gorini}
Vittorio Gorini, Andrzej Kossakowski, E.~C.~G. Sudarshan, \enquote{Completely
  positive dynamical semigroups of N‐level systems},
  \href{http://dx.doi.org/10.1063/1.522979}{\emph{Journal of Mathematical
  Physics}}, \textbf{17}: 5 (1976), 821--825,
  \eprint{https://aip.scitation.org/doi/pdf/10.1063/1.522979}.

\bibitem{Lindblant}
Frederik Nathan, Mark~S. Rudner, \enquote{Universal Lindblad equation for open
  quantum systems},
  \href{http://dx.doi.org/10.1103/PhysRevB.102.115109}{\emph{Phys. Rev. B}},
  \textbf{102} (2020), 115109.

\bibitem{Tarasov}
Vasily Tarasov, \enquote{Quantum Mechanics of Non-Hamiltonian and Dissipative
  Systems}, Elsevier. Monograph Series On Nonlinear Science and Complexity.
  ISBN: 9780080559711, {(2008)}.

\bibitem{Kubo1}
Ryogo Kubo, \enquote{Statistical-Mechanical Theory of Irreversible Processes.
  I. General Theory and Simple Applications to Magnetic and Conduction
  Problems}, \href{http://dx.doi.org/10.1143/JPSJ.12.570}{\emph{Journal of the
  Physical Society of Japan}}, \textbf{12}: 6 (1957), 570--586,
  \eprint{https://doi.org/10.1143/JPSJ.12.570}.

\bibitem{Kubo2}
Ryogo Kubo, Mario Yokota, Sadao Nakajima, \enquote{Statistical-Mechanical
  Theory of Irreversible Processes. II. Response to Thermal Disturbance},
  \href{http://dx.doi.org/10.1143/JPSJ.12.1203}{\emph{Journal of the Physical
  Society of Japan}}, \textbf{12}: 11 (1957), 1203--1211,
  \eprint{https://doi.org/10.1143/JPSJ.12.1203}.

\bibitem{Mouas}
Mohamed Mouas, Jean-Georges Gasser, Slimane Hellal, Benoit Grosdidier, Ahmed
  Makradi, Salim Belouettar, \enquote{Diffusion and viscosity of liquid tin:
  Green-Kubo relationship-based calculations from molecular dynamics
  simulations}, \href{http://dx.doi.org/10.1063/1.3687243}{\emph{The Journal of
  chemical physics}}, \textbf{136} (2012), 094501.

\bibitem{Liu}
Pu~Liu, Edward Harder, B.~J. Berne,
  \href{http://dx.doi.org/10.1021/jp0375057}{\emph{The Journal of Physical
  Chemistry B}}, \textbf{108}: 21 (2004), 6595--6602,
  \eprint{https://doi.org/10.1021/jp0375057}.

\bibitem{Volkov}
NA~Volkov, MV~Posysoev, AK~Shchekin, \enquote{The effect of simulation cell
  size on the diffusion coefficient of an ionic surfactant aggregate},
  \emph{Colloid Journal}, \textbf{80}: 3 (2018), 248--254.

\bibitem{AdzhemyanKuni}
L.~Ts. Adzhemyan, F.~M. Kuni, T.~Yu. Novozhilova, \enquote{Nonlinear
  generalization of Mori's method of projection operators},
  \href{http://dx.doi.org/10.1007/BF01035649}{\emph{Theoretical and
  Mathematical Physics}}, \textbf{18} (1974), 383–392.

\bibitem{Eliashberg}
G.~M. Eliashberg, \enquote{Transport equation for a degenerate system of fermi
  particles}, \emph{Sov. Phys. JETP}, \textbf{14} (1962), 886--892.

\bibitem{KelRev2}
P~I Arseev, \enquote{On the nonequilibrium diagram technique: derivation, some
  features, and applications},
  \href{http://dx.doi.org/10.3367/ufne.0185.201512b.1271}{\emph{Physics-Uspekhi}},
  \textbf{58}: 12 (2015), 1159--1205.

\bibitem{Sieberer}
L.~M. Sieberer, A.~Chiocchetta, A.~Gambassi, U.~C. T\"auber, S.~Diehl,
  \enquote{Thermodynamic equilibrium as a symmetry of the Schwinger-Keldysh
  action}, \href{http://dx.doi.org/10.1103/PhysRevB.92.134307}{\emph{Phys. Rev.
  B}}, \textbf{92} (2015), 134307.

\bibitem{Vasiljev2}
A.~N. Vasilev, \enquote{The field theoretic renormalization group in critical
  behavior theory and stochastic dynamics}, Chapman \& Hall/CRC, Boca Raton,
  FL, 2004, \doi{10.1201/9780203483565}, translated from the 1998 Russian
  original by Patricia A. de Forerand-Millard and revised by the author,
  \urlprefix\url{https://doi.org/10.1201/9780203483565}.

\bibitem{ELSM}
Juha Honkonen, M.V. Komarova, Yu.G. Molotkov, M.Yu. Nalimov, \enquote{Effective
  large-scale model of boson gas from microscopic theory},
  \href{http://dx.doi.org/https://doi.org/10.1016/j.nuclphysb.2018.12.015}{\emph{Nuclear
  Physics B}}, \textbf{939} (2019), 105--129.

\bibitem{Maghrebi}
Mohammad~F. Maghrebi, Alexey~V. Gorshkov, \enquote{Nonequilibrium many-body
  steady states via Keldysh formalism},
  \href{http://dx.doi.org/10.1103/PhysRevB.93.014307}{\emph{Phys. Rev. B}},
  \textbf{93} (2016), 014307.

\bibitem{KelRev1}
Robert van Leeuwen, Nils Dahlen, Gianluca Stefanucci, Carl-Olof Almbladh,
  \enquote{Introduction to the Keldysh formalism and applications to
  time-dependent density-functional theory}, \emph{Lecture Notes in Physics,
  Springer, Berlin}, \textbf{706}.

\bibitem{KelRev3}
Felix Haehl, R.~Loganayagam, Mukund Rangamani, \enquote{Schwinger-Keldysh
  formalism I: BRST symmetries and superspace},
  \href{http://dx.doi.org/10.1007/JHEP06(2017)069}{\emph{Journal of High Energy
  Physics}}, \textbf{2017} (2017), 69.

\bibitem{KelRev4}
Marcelo Janovitch~B. Pereira, \enquote{Keldysh Field Theory}, \emph{Lecture
  notes, University of São Paulo (2019)}.

\bibitem{KelRev5}
Michael Geracie, Felix~M. Haehl, R.~Loganayagam, Prithvi Narayan, David~M.
  Ramirez, Mukund Rangamani, \enquote{Schwinger-Keldysh superspace in quantum
  mechanics}, \href{http://dx.doi.org/10.1103/PhysRevD.97.105023}{\emph{Phys.
  Rev. D}}, \textbf{97} (2018), 105023.

\bibitem{KelRev6}
Alex Kamenev, Alex Levchenko, \enquote{Keldysh technique and non-linear
  sigma-model: basic principles and applications},
  \href{http://dx.doi.org/10.1080/00018730902850504}{\emph{Advances in
  Physics}}, \textbf{58}: 3 (2009), 197--319,
  \eprint{https://doi.org/10.1080/00018730902850504}.

\bibitem{KelRev7}
Jørgen Rammer, \enquote{Quantum Field Theory of Non-Equilibrium States,
  Cambridge University Press}, 2007, \doi{10.1017/CBO9780511618956},
  \urlprefix\url{https://doi.org/10.1017/CBO9780511618956}.

\bibitem{KelRev8}
J.~Rammer, H.~Smith, \enquote{Quantum field-theoretical methods in transport
  theory of metals},
  \href{http://dx.doi.org/10.1103/RevModPhys.58.323}{\emph{Rev. Mod. Phys.}},
  \textbf{58} (1986), 323--359.

\bibitem{K-B}
L.~P. Kadanoff, G.~Baym, \enquote{Quantum Statistical Mechanics}, W.A. Benjamin
  Inc., New York, 1962.

\bibitem{Tanaka}
Tomohiro Tanaka, Yusuke Nishida, \enquote{Thermal conductivity of a weakly
  interacting Bose gas by quasi-one dimensionality}, \emph{preprint
  arXiv:2203.04936}.

\bibitem{Sangyong}
Sangyong Jeon, \enquote{Hydrodynamic transport coefficients in relativistic
  scalar field theory},
  \href{http://dx.doi.org/10.1103/PhysRevD.52.3591}{\emph{Phys. Rev. D}},
  \textbf{52} (1995), 3591--3642.

\bibitem{Sangyong2}
Sangyong Jeon, Laurence~G. Yaffe, \enquote{From quantum field theory to
  hydrodynamics: Transport coefficients and effective kinetic theory},
  \href{http://dx.doi.org/10.1103/PhysRevD.53.5799}{\emph{Phys. Rev. D}},
  \textbf{53} (1996), 5799--5809.

\bibitem{Yoshimasa}
Yoshimasa Hidaka, Teiji Kunihiro, \enquote{Renormalized linear kinetic theory
  as derived from quantum field theory: A novel diagrammatic method for
  computing transport coefficients},
  \href{http://dx.doi.org/10.1103/PhysRevD.83.076004}{\emph{Phys. Rev. D}},
  \textbf{83} (2011), 076004.

\bibitem{Nieves}
Jose Nieves, Sarira Sahu, \enquote{Taming the pinch singularities in the
  two-loop neutrino self-energy in a medium}, \emph{preprint arXiv:2104.04459}.

\bibitem{Akhmedov1}
E.~T. Akhmedov, U.~Moschella, F.~K. Popov, \enquote{Characters of different
  secular effects in various patches of de Sitter space},
  \href{http://dx.doi.org/10.1103/PhysRevD.99.086009}{\emph{Phys. Rev. D}},
  \textbf{99} (2019), 086009.

\bibitem{Akhmedov2}
E.~T. Akhmedov, Ph. Burda, \enquote{Solution of the Dyson-Schwinger equation on
  a de Sitter background in the infrared limit},
  \href{http://dx.doi.org/10.1103/PhysRevD.86.044031}{\emph{Phys. Rev. D}},
  \textbf{86} (2012), 044031.

\bibitem{Akhmedov3}
E.~T. Akhmedov, F.~K. Popov, V.~M. Slepukhin, \enquote{{Infrared dynamics of
  the massive \ensuremath{\phi}4 theory on de Sitter space}},
  \href{http://dx.doi.org/10.1103/PhysRevD.88.024021}{\emph{Phys. Rev. D}},
  \textbf{88} (2013), 024021, \eprint{1303.1068}.

\bibitem{Akhmedov4}
E.~Akhmedov, Nikita Astrakhantsev, Fedor Popov, \enquote{Secularly growing loop
  corrections in strong electric fields},
  \href{http://dx.doi.org/10.1007/JHEP09(2014)071}{\emph{Journal of High Energy
  Physics}}, \textbf{2014}.

\bibitem{Abrikosov}
A.A. Abrikosov, L.P. Gorkov, I.E. Dzyaloshinski, R.A. Silverman,
  \enquote{Methods of Quantum Field Theory in Statistical Physics}, Dover Books
  on Physics, Dover Publications, 2012,
  \urlprefix\url{https://books.google.ru/books?id=JYTCAgAAQBAJ}.

\bibitem{A}
Jens~O Andersen, \enquote{{Theory of the weakly interacting Bose gas}},
  \href{http://dx.doi.org/10.1103/RevModPhys.76.599}{\emph{Rev. Mod. Phys.}},
  \textbf{76} (2004), 599, \eprint{cond-mat/0305138}.

\bibitem{KUFG}
J.~Honkonen, \enquote{Contour-ordered Green’s functions in stochastic field
  theory}, \href{http://dx.doi.org/10.1007/s11232-013-0069-2}{\emph{Theoret.
  and Math. Phys.}}, \textbf{175} (2013), 827–834.

\bibitem{osnovnaya}
J.~Honkonen, M.~V. Komarova, Yu.~G. Molotkov, M.~Yu. Nalimov, \enquote{Kinetic
  Theory of Boson Gas JO - Theoretical and Mathematica},
  \href{http://dx.doi.org/10.1134/S0040577919090095}{\emph{Theoret. and Math.
  Phys.}}, \textbf{200} (2019), 1360--1373.

\bibitem{Vasiljev1}
A.~N. Vasil'ev, \enquote{{Functional Methods in Quantum Field Theory and
  Statistical Physics}}, Gordon and Breach, Amsterdam, 1998.

\bibitem{N}
Yu.~A. Zhavoronkov, M.~V. Komarova, Yu.~G. Molotkov, M.~Yu. Nalimov,
  \enquote{Critical Dynamics of the Phase Transition to the Superfluid State},
  \href{http://dx.doi.org/10.1134/S0040577919080142}{\emph{Theoret. and Math.
  Phys.}}, \textbf{200} (2019), 1237--1251.

\bibitem{K}
L.~V. Keldysh, \enquote{Diagram technique for nonequilibrium processes},
  \emph{Sov.~Phys.~JETP}, \textbf{20} (1965), 1018, [Zh. Eksp. Theor. Fiz.\
  {\bf 47}, 1515 (1964)].

\bibitem{S}
Julian Schwinger, \enquote{Brownian Motion of a Quantum Oscillator},
  \href{http://dx.doi.org/10.1063/1.1703727}{\emph{Journal of Mathematical
  Physics}}, \textbf{2}: 3 (1961), 407--432.

\bibitem{Markku22}
M.~J. Hyrk\"as, D.~Karlsson, R.~van Leeuwen, \enquote{Cutting rules and
  positivity in finite temperature many-body theory}, \emph{preprint
  arXiv:2203.11083}.

\bibitem{Hua}
Rudolph~C Hwa, Vigdor~L Teplitz, \enquote{{Homology and Feynman integrals}},
  Mathematical physics monograph series, Benjamin, New York, NY, 1966,
  \urlprefix\url{http://cds.cern.ch/record/102287}.

\bibitem{Wyld}
H.W Wyld, \enquote{Formulation of the theory of turbulence in an incompressible
  fluid},
  \href{http://dx.doi.org/https://doi.org/10.1016/0003-4916(61)90056-2}{\emph{Annals
  of Physics}}, \textbf{14} (1961), 143--165.

\bibitem{Landau}
L.~D. Landau, E.~M. Lifshitz, \enquote{Fluid Mechanics, Second Edition: Volume
  6 (Course of Theoretical Physics)}, Course of theoretical physics / by L. D.
  Landau and E. M. Lifshitz, Vol. 6, Butterworth-Heinemann, 2 edn., 1987,
  \urlprefix\url{http://www.worldcat.org/isbn/0750627670}.

\bibitem{5loops}
L.Ts. Adzhemyan, D.A. Evdokimov, M.~Hnatič, et~al., \enquote{Model A of
  critical dynamics: 5-loop epsilon expansion study},
  \href{http://dx.doi.org/https://doi.org/10.1016/j.physa.2022.127530}{\emph{Physica
  A: Statistical Mechanics and its Applications}}, \textbf{600} (2022), 127530.

\bibitem{5loops2}
L.~Ts. Adzhemyan, D.~A. Evdokimov, M.~Hnati\v{c}, et~al., \enquote{{The dynamic
  critical exponent z for 2d and 3d Ising models from five-loop
  \ensuremath{\varepsilon} expansion}},
  \href{http://dx.doi.org/10.1016/j.physleta.2021.127870}{\emph{Phys. Lett.
  A}}, \textbf{425} (2022), 127870, \eprint{2111.04719}.

\bibitem{tauber}
Uwe~C T{\"a}uber, \enquote{Critical Dynamics: A Field Theory Approach to
  Equilibrium and Non-Equilibrium Scaling Behavior}, Cambridge University
  Press, 2014.

\bibitem{tauber2}
Uwe~C. Täuber, \enquote{Renormalization Group: Applications in Statistical
  Physics},
  \href{http://dx.doi.org/https://doi.org/10.1016/j.nuclphysbps.2012.06.002}{\emph{Nuclear
  Physics B - Proceedings Supplements}}, \textbf{228} (2012), 7--34, “Physics
  at all scales: The Renormalization Group” Proceedings of the 49th
  Internationale Universitätswochen für Theoretische Physik.

\bibitem{Fmodel}
M.~Gnatich, M.~V. Komarova, M.~Yu. Nalimov, \enquote{Microscopic justification
  of the stochastic f-model of critical dynamics},
  \href{http://dx.doi.org/10.1007/s11232-013-0064-7}{\emph{Theoretical and
  Mathematical Physics}}, \textbf{175}: 3 (2013), 131--142.

\bibitem{Ershov1}
A.~M. Il’in, A.~A. Ershov, \enquote{Asymptotics of two-dimensional integrals
  depending singularly on a small parameter},
  \href{http://dx.doi.org/10.1134/S008154381005010X}{\emph{Proceedings of the
  Steklov Institute of Mathematics}}, \textbf{268} (2010), 131--142.

\bibitem{Ershov2}
A.~A. Ershov, M.~I Rusanova, \enquote{Asymptotics of multidimensional integrals
  depending singularly on a small parameter},
  \href{http://dx.doi.org/10.1134/S008154381705008X}{\emph{Proceedings of the
  Steklov Institute of Mathematics}}, \textbf{297} (2017), 72--80.

\bibitem{Ilin}
Danilin A.~R. Il'in, A.~M., \enquote{{Asymptotic methods in analysis (In
  Russian)}}, Moscow, Fizmatlit Publ, 2009.

\bibitem{Mattuk}
R.~Mattuck, \enquote{{A Guide to Feynman Diagrams in the Many-Body Problem:
  Second Edition}}, Courier Corporation, 2012.

\bibitem{Graphene19}
Joseph Sulpizio, Lior Ella, Asaf Rozen, et~al., \enquote{Visualizing Poiseuille
  flow of hydrodynamic electrons},
  \href{http://dx.doi.org/10.1038/s41586-019-1788-9}{\emph{Nature}},
  \textbf{576} (2019), 75--79.

\bibitem{Novoselov05}
K.~Novoselov, A.K. Geim, S.~Morozov, et~al., \enquote{Two-Dimensional Gas of
  Massless Dirac Fermions in Graphene},
  \href{http://dx.doi.org/10.1038/nature04233}{\emph{Nature}}, \textbf{438}
  (2005), 197--200.

\end{thebibliography}
\end{document}